\documentclass[11pt]{article}

\usepackage{epsfig}

\newcommand\nc{\newcommand}

\nc{\textsize}[2]{
\textwidth=#1
\oddsidemargin=8.5in
\addtolength\oddsidemargin{-2in}
\addtolength\oddsidemargin{-\textwidth}
\divide\oddsidemargin by2
\evensidemargin=\oddsidemargin
\textheight=#2
\topmargin=11in
\addtolength\topmargin{-3in}
\addtolength\topmargin{-\textheight}
\divide\topmargin by2
}

\newtheorem{theorem}{Theorem}[section]
\newtheorem{lemma}[theorem]{Lemma}
\newtheorem{cor}[theorem]{Corollary}

\newtheorem{ques}[theorem]{Question}
\newtheorem{prob}[theorem]{Problem}

\nc{\crl}[2]{\begin{cor}\label{crl:#1} #2 \end{cor}}
\nc{\dfn}[2]{\begin{defn}\label{def:#1} #2 \end{defn}}
\nc{\lem}[2]{\begin{lemma}\label{lem:#1} #2 \end{lemma}}
\nc{\thm}[2]{\begin{theorem}\label{thm:#1} #2 \end{theorem}}
\nc{\que}[2]{\begin{ques}\label{que:#1} #2 \end{ques}}
\nc{\pro}[2]{\begin{prob}\label{pro:#1} #2 \end{prob}}

\nc{\fig}[4]{\begin{figure}[hbt]
\centerline{
\epsfysize=#2in
\epsffile{#4}
}
\caption{#3}
\label{fig:#1}
\end{figure}}

\nc{\refc}[1]{Corollary~\ref{crl:#1}}
\nc{\refd}[1]{Definition~\ref{def:#1}}
\nc{\reff}[1]{Figure~\ref{fig:#1}}
\nc{\refl}[1]{Lemma~\ref{lem:#1}}
\nc{\reft}[1]{Theorem~\ref{thm:#1}}
\nc{\refb}[1]{Problem~\ref{pro:#1}}

\def\qed{\rule{1.5mm}{3mm}}
\def\es{\emptyset}

\nc{\proof}[1]{ {\bf Proof.} #1 \hfill \qed\par}

\nc{\pf}[1]{{\bf Proof.} #1 \hfill \qed\par}

\textsize{5.4in}{8.5in}

\title{\Large \bf Vertex Covers Revisited: \\ Indirect Certificates and FPT Algorithms}
\author{{\large CAI Leizhen\thanks{Partially supported by CUHK Direct Grant 4055069}}  \\ \\
\normalsize \sl Department of Computer Science and Engineering \\
\normalsize \sl  The Chinese University of Hong Kong \\
\normalsize \sl Shatin, N.T., Hong Kong SAR, China \\
\normalsize \tt E-mail:lcai@cse.cuhk.edu.hk \\}
\date{July 27, 2018}

\begin{document}
\maketitle

\begin{abstract}
The classical NP-complete problem {\sc Vertex Cover} requires us to determine
whether a graph contains at most $k$ vertices that cover all edges.
In spite of its intractability, the problem can be solved in FPT time for
parameter $k$ by various techniques.

In this paper, we present half a dozen new and simple FPT algorithms for {\sc Vertex Cover}
with respect to parameter $k$.
For this purpose, we explore structural properties of vertex covers and use
these properties to obtain FPT algorithms by iterative compression, colour coding,
and indirect certificating methods.
In particular, we show that every graph with a $k$-vertex cover admits an indirect
certificate with at most $k/3$ vertices, which lays the foundation of three 
new FPT algorithms based on random partition and random selection.

\vspace{0.5cm}


\end{abstract}

\newpage

\section{Introduction}

We start with the following extremely simple algorithm for a graph $G$,
where $N(M)$ denotes the open neighbourhood of marked vertices, and
a graph in \reff{single-vertex}:

\vspace{0.5cm}
\fbox{\begin{minipage}[c]{0.9\textwidth}
Randomly mark each vertex with probability 1/2 and output $N(M)$.
\end{minipage}}
\vspace{0.5cm}

What problem does the above algorithm solve?  Surprisingly, we will show later that 
it actually finds, with probability at least $4^{-k}$, a vertex cover consisting of at most $k$ vertices, 
if $G$ admits such a vertex cover.

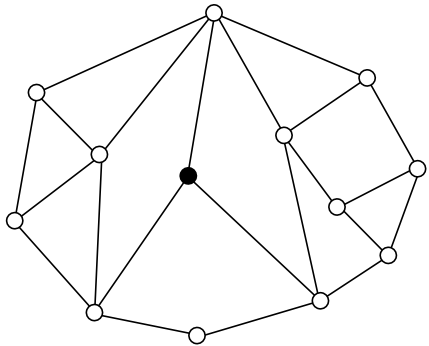
\fig{single-vertex}{1.8}{A graph with one marked vertex as certificate.}{single-vertex.fig}

As for the graph in \reff{single-vertex}, the single dark vertex in the graph,
instead of a vertex cover with eight vertices, can be used 
as an indirect certificate to certify that the graph has such a vertex cover.
And the existence of such small indirect certificates has an intimate connection 
with the above algorithm.

Yes, the subject of this paper is the classical {\sc Vertex Cover} problem
and we are interested in designing new FPT algorithms for the problem.
\begin{quote}
{\sc Vertex Cover}\\
{\sc Instance}: Graph $G = (V, E)$, and positive integer $k$ as parameter. \\
{\sc Question}: Does $G$ contain a vertex cover of size at most $k$, i.e.,
at most $k$ vertices that cover all edges?
\end{quote}

Our motivations are multifold though not for the race of fastest FPT algorithms:
\begin{itemize}
\item We wish to find structural properties that give us a better understanding of the problem
and will be useful for obtaining FPT algorithms.
\item Simple FPT algorithms of different flavours for {\sc Vertex Cover} are
instrumental in teaching FPT algorithms as the problem is almost always used
as the first example in such an endeavour.
\item It is also intellectually challenging to find different methods 
to solve a problem, aiming for elegant and simple algorithms and proofs.
\end{itemize}

There are about a dozen different FPT algorithms for 
{\sc Vertex Cover}~\cite{Abu,Buss,Chen,Dehne,Fellows,Kurt}, 
and the current fastest one of Chen, Kanj, and Xia~\cite{Chen}
runs in $O(1.2738^k + kn)$ time.
Most such algorithms are based on structural properties of vertex covers.
For instance, the basic bounded search tree algorithm~\cite{Kurt, Fellows} 
is based on the trivial fact that we need to choose at least one end 
of an edge to cover the edge, 
and the kernelization algorithm of Buss~\cite{Buss} uses the simple observation
that any vertex of degree more than $k$ is forced to be in any vertex cover 
of size $k$.

In this paper, we present half a dozen new and simple FPT algorithms for 
{\sc Vertex Cover}:
\begin{itemize}
\item Three FPT algorithms based on the idea of ``indirect certificating'' that explores 
indirect certificates for NP-complete problems and generalizes the random separation 
method of Cai, Chan and Chan~\cite{CCC} (\S2).
The foundation of these three algorithms relies on the property that every yes-instance 
of {\sc Vertex Cover} admits an indirect certificate of size at most $k/3$.
\item FPT algorithms that rely on characterizations of minimum vertex covers 
and use the iterative compression method of Reed, Smith and Vetta~\cite{Reed}(\S3).
\item A linear-time algorithm for finding colourful vertex covers, which yields 
an FPT algorithm by the colour coding method of Alon, Yuster, and Zwick~\cite{Alon}(\S4).
\end{itemize}

In this paper, $G = (V, E)$ is a graph with $m$ edges and $n$ vertices.
For convenience, we use {\em $k$-vertex cover} for any vertex cover 
with at most $k$ vertices.
Note that $m \le kn$ for any graph containing a $k$-vertex cover.

For any subset $S$ of vertices, $N(S)$ denotes the {\em open neighbourhood} of $S$,
i.e., all vertices in $V - S$ that are adjacent to $S$, 
and $N[S]$ the {\em closed neighbourhood} of $S$ which equals $S \cup N(S)$.
By {\em indirect certificate}, we refer to a structure $\chi$ disjoint from a
solution $X$ such that $\chi$ can be used to certify the existence of $X$
in polynomial time, and we say that $\chi$ is a {\em small indirect certificate}
if its size is bounded above by a function of $|X|$ alone.
For two disjoint subsets $\chi$ and $X$ of vertices $V$, a partition $(V', V - V')$ of $V$ is
a {\em valid $(\chi, X)$-partition} if $\chi \subseteq V'$ and $X \subseteq V - V'$.

\section{Indirect certificates}

The random separation method of Cai, Chan and Chan~\cite{CCC} is an innovative 
method for designing FPT algorithms to solve graph problems.
The basic idea of their method is to randomly partition the vertex set of a graph
into red and blue vertices to separate a {\em $k$-solution} $X$ (i.e., a solution with 
$\le k$ vertices) into blue components, 
and then choose appropriate blue components to form a $k$-solution.
The method is very effective for problems on degree-bounded graphs, and basically can be 
used to obtain FPT algorithms for any problems dealing with ``local properties''.
Unfortunately, the original idea of their random separation method requires the size
of $N[X]$ be bounded by a function of $k$, which limits the power of their
method to basically degree-bounded graphs.
And it seems impossible to obtain an FPT algorithm for {\sc Vertex Cover}
by random separation.

However, a close examination of the method reveals that it is a special case 
of a more general idea of {\em indirect certificating} that uses 
small indirect certificates as helpers to find $k$-solutions:
\begin{quote}
{\em First discover a small-size structure $\chi$ that is disjoint from
a $k$-solution $X$ and can be used as a certificate for a yes-instance.
Then use random partition to produce a valid $(\chi, X)$-partition,
and find a $k$-solution with the aid of $\chi$.}
\end{quote}
In fact two examples in Cai, Chan and Chan~\cite{CCC}, including the problem of 
finding a maximum-weight independent set of size $k$ in planar graphs, implicitly 
used this general idea.

This general idea of indirect certificating seems quite potential for obtaining FPT algorithms, 
and there have been a few FPT algorithms based on this idea: 
Cygan et al.~\cite{Cygan} have designed an FPT algorithm for obtaining an Eulerian graph
by deleting at most $k$ edges, and Cai and Ye~\cite{CaiYe} have presented FPT algorithms 
for finding two edge-disjoint $(s,t)$-paths with some length constraints.

In the following three subsections, we will use the idea of indirect certificating to design three
randomized FPT algorithms.
Foundations of these algorithms are laid by
a simple indirect certificate of size $k$ (\refl{vc-cert}) and 
an elaborated indirect certificate of size $k/3$ (\reft{small-cert}) for 
{\sc Vertex Cover}.

\subsection{Certificates with $k$ vertices}

We will show that the algorithm that starts the paper is a randomized FPT 
algorithm for {\sc Vertex Cover} based on the idea of indirect certificating.
First we present an alternative certificate $\chi$ for the problem,
as we can use $\chi$, instead of a $k$-vertex cover, to verify that a graph is
indeed a yes-instance of {\sc Vertex Cover}.

\lem{vc-cert}{
For any minimal vertex cover $X$ of a graph $G = (V,E)$, 
$G$ contains an independent set $\chi$ with at most $|X|$ vertices such that $N(\chi) = X$.
}
\pf{Since $X$ is a minimal vertex cover, 
every vertex in $X$ covers at least one edge in cut $[X, V - X]$.
For each vertex $v \in X$, arbitrarily choose an adjacent vertex $v' \in V - X$
and let $\chi = \{v' : \mbox{ $v \in X$ } \}$.
Then $\chi$ clearly has the properties in the lemma as $V - X$ is an independent set.
}

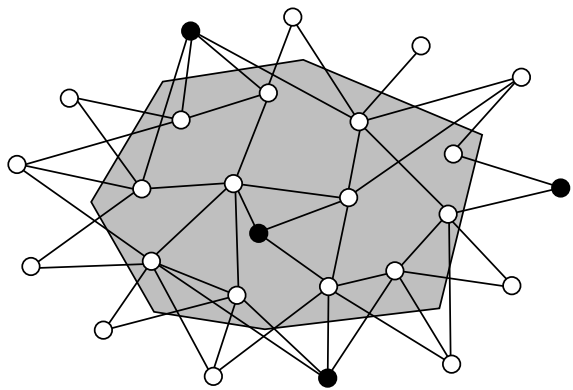
\fig{vc}{2}{Vertices inside the shaded polygon form a vertex cover $X$ of the graph,
and dark vertices $\chi$ yield an indirect certificate of size 4.
Note that $X$ is not a minimal vertex cover, and $N(\chi)$ gives a smaller one.}{smallcert.fig}

The above simple structural property of vertex cover is surprisingly
useful for designing a randomized FPT algorithm for {\sc Vertex Cover}.
Let $G = (V, E)$ be a graph that admits a $k$-vertex cover $X$, 
and we may assume that $X$ is a minimal vertex cover.
Then by \refl{vc-cert}, $V - X$ contains at most $k$ vertices $\chi$ disjoint from $X$,
and $\chi$ forms an indirect certificate that can be used to verify the existence of $X$.
As $X$ and $\chi$ reside in two disjoint sets, random partition becomes
a natural tool to obtain a randomized FPT algorithm for {\sc Vertex Cover}.

We randomly and independently colour each vertex by either red or blue with equal probability 
to form a random partition $\{V_r, V_b\}$ of vertices of $G$,
where $V_r$ and $V_b$ respectively are red and blue vertices.
There are two key points to note:
\begin{itemize}
\item A random red-blue colouring has probability at least $4^{-k}$ to produce a valid
$(\chi, X)$-partition.
\item Once we have a valid $(\chi, X)$-partition, the neighbourhood of red vertices 
yields a required vertex cover.
\end{itemize}

The following is the algorithm at the beginning of the paper
in the language of red-blue colouring.

\vspace{0.5cm}
\fbox{\begin{minipage}[c]{0.9\textwidth}
\noindent Algorithm VC-IC[$k$]\\
\noindent Input: A graph $G = (V,E)$. \\
\noindent Output: A $k$-vertex cover $X$ of $G$, if it exists.

\begin{enumerate}
\item Randomly and independently colour each vertex red or blue 
with probability 1/2 to generate a random vertex-partition $(V_r,V_b)$ of $G$.

\item Output $N(V_r)$ as $X$.
\end{enumerate}
\end{minipage}}
\vspace{0.5cm}

\thm{vc-k}{{\rm Algorithm VC-IC[$k$]} finds, with probability at least $4^{-k}$,
a $k$-vertex cover of $G$, if it exists, in $O(kn)$ time,
and therefore solves {\sc Vertex Cover} in $O(4^kkn)$ expected time.
}
\pf{
Let $X$ be a minimal $k$-vertex cover of $G$.
By \refl{vc-cert}, $G$ contains at most $k$ vertices $\chi$
such that $N(\chi) = X$.
Then with probability at least $2^{-k}$, Step 1 colours all vertices in $\chi$ red
and all vertices in $X$ blue.
It follows that Step 1 produces a valid $(\chi,X)$-partition $(V_r,V_b)$ 
with probability at least $4^{-k}$, 
and therefore the expected number of random red-blue colourings required  
to produce a valid $(\chi,X)$-partition is $O(4^k)$.

For any valid $(\chi, X)$-partition $(V_r,V_b)$ of vertices, we see that 
$N(V_r)$ contains $X$ as all vertices of $\chi$ are red and $N(\chi) = X$ by \refl{vc-cert}.
On the other hand, $V_r \subseteq V - X$ and hence $N(V_r)$ contains no vertex of $V - X$
as $V_r$ is an independent set.
Therefore $N(V_r) = X$ and the algorithm correctly returns $X$ for such a partition
in $O(kn)$ time as $m \le kn$.
}

The algorithm can be made into a deterministic FPT algorithm with running time 
$4^kk^{O(\log k)}n\log n$ by a family of $(n, 2k)$-universal sets for
derandomization~\cite{NSS}.

\subsection{Smaller certificates}

It is perhaps surprising that the mere existence of an indirect certificate $\chi$ 
of size $k$ enables us to find a $k$-vertex cover $X$ in expected FPT time,
even without knowing the actual $\chi$.
This motivates us to search for smaller indirect
certificates, and indeed we can significantly reduce the size of certificates to 
$k/3$ only, which leads to an improvement of our FPT algorithm.

We start with a general property regarding a relatively small number
of vertices in connection with minimum vertex covers to lay the foundation of
certificates of size $k/3$ for {\sc Vertex Cover}.
For any integer $d \ge 0$, let $V_d$ denote the set of vertices of degree at least $d$.

\lem{smaller-cert}{
Every graph $G$ with a $k$-vertex cover contains at most $k/d$ vertices $\chi$, where $d$ is 
any positive integer, such that for every minimum vertex cover $X^*$ of 
$G - (N[\chi] \cup V_d(G - N[\chi]))$,
vertices $N(\chi) \cup V_d(G - N[\chi]) \cup X^*$ is a minimum vertex cover of $G$.
}
\pf{
Let $X$ be a minimum vertex cover of $G$.
We construct $\chi$ by choosing at most $k/3$ vertices from $V - X$
as follows. Note that all vertices of $X$ are unmarked initially, and we process
vertices $V - X$ in an arbitrary given order. 
Also note that for any vertex $v \in V - X$, $N(v) \subseteq X$.

\begin{quote}
For each $v \in V - X$, we put $v$ into $\chi$ and mark all unmarked vertices in $N(v)$
whenever $v$ is adjacent to at least $d$ unmarked vertices in $X$.
\end{quote}

For convenience let $H = G - N[\chi]$ and $G^* = H - V_d(H)$, and note
that $X^*$ is a minimum vertex cover of $G^*$.
It is clear that $|\chi| \le k/d$ as we mark at least $d$ unmarked 
vertices in $X$ each time we put a vertex into $\chi$.
We show that vertices $N(\chi) \cup V_d(H) \cup X^*$ is a minimum vertex cover of $G$.
First we note that $V_d(H) \subseteq X$ as all vertices of $V - X$ in $H$ have 
degree less than $d$ by the choice of $\chi$.
Also, $N(\chi) \cup V_d(H) \cup X^*$ is clearly a vertex cover of $G$
as all edges outside $G^*$ are covered by  $N(\chi) \cup V_d(H)$.

It remains to be shown that $N(\chi) \cup V_d(H) \cup X^*$ has the same number of vertices as $X$.
It is helpful to note that $X$ consists of three disjoint parts: $N(\chi)$, $V_d(H)$,
and the remaining vertices $R$.
Since $V_d(H) \cup R$ is a minimum vertex cover of $H$, $R$ is a minimum vertex cover of $G^*$.
Therefore $|R| = |X^*|$ as $X^*$ is also a minimum vertex cover of $G^*$. 
It follows that $|N(\chi) \cup V_d(H) \cup X^*| = |X|$ and 
hence $N(\chi) \cup V_d(H) \cup X^*$ is a minimum vertex cover of $G$.
}

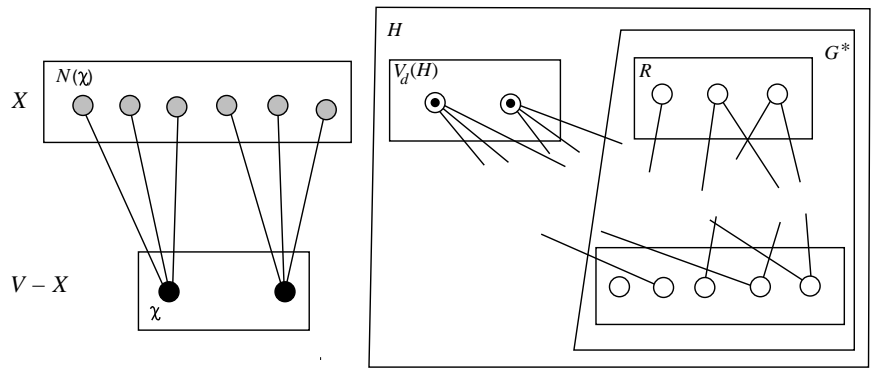
\fig{third}{1.9}{The structure of $G$ with respect to vertex cover $X$ and indirect certificate
$\chi$.}{smaller.fig}

The above lemma has interesting and surprising consequences.
For {\sc Vertex Cover}, \refl{smaller-cert} for $d = k+1$ actually gives us an
alternative perspective  of the kernelization algorithm of Buss~\cite{Buss}: 
In this case, both $\chi$ and $N[\chi]$ are empty, and
hence $V_{k+1}(G)$ are vertices forced to be in any $k$-vertex cover, and graph 
$G - V_{k+1}(G)$ has maximum degree at most $k$ and forms a kernel of the problem.

Turing to indirect certificates, we recall that {\sc Vertex Cover} remains NP-complete for
cubic graphs but is trivially solved in linear time for graphs of maximum degree at most two,
i.e., graphs consisting of disjoint union of paths and cycles.
By setting $d = 3$ in \refl{smaller-cert}, we see that $k/3$ vertices are sufficient to
certify yes-instances of {\sc Vertex Cover}, which yields the single-vertex certificate 
in \reff{single-vertex} at the beginning of the paper.

\thm{small-cert}{
Every yes-instance $(G, k)$ of {\sc Vertex Cover} admits a certificate $\chi$ 
with at most $k/3$ vertices.
}
\pf{
Let $X$ be a minimum vertex cover of $G$. 
Then $|X| \le k$ and we use $\chi$ in \refl{smaller-cert} with $d = 3$ as a certificate
for our verification algorithm.

Let $G^* = G - (N[\chi] \cup V_3(G - N[\chi]))$.
Given $G$ and $\chi$ as input, we first compute a minimum vertex cover $X^*$ of $G^*$,
which by \refl{smaller-cert} yields $N(\chi) \cup V_3(G - N[\chi]) \cup X^*$ as a minimum vertex cover of $G$.
Since $G^*$ is a graph of maximum degree two, the computation of $X^*$ takes linear time.
Therefore we can verify that $(G, k)$ is indeed a yes-instance of {\sc Vertex Cover} 
in polynomial time.
}

With \refl{smaller-cert} at hand, we can easily improve the success probability of 
Algorithm VC-IC$[k]$ from $4^{-k}$ to $2^{-\frac{4}{3}k} > 2.5199^{-k}$.
In fact, we can further improve it to $2.1166^{-k}$ by optimizing the
probabilities of colouring vertices red or blue. 

\vspace{0.5cm}
\fbox{\begin{minipage}[c]{0.9\textwidth}
\noindent Algorithm VC-IC[$k/3$]\\
\noindent Input: A graph $G = (V,E)$.\\
\noindent Output: A $k$-vertex cover $X$ of $G$, if it exists.

\begin{enumerate}
\item Randomly and independently colour each vertex red with probability 
1/4 and blue with probability 3/4 to generate a random vertex-partition $(V_r,V_b)$ of $G$.

\item Compute a minimum vertex cover $X^*$ of $G - (N[V_r] \cup V_3(G - N[V_r])$, 
and output $N(V_r) \cup V_3(G - N[V_r]) \cup X^*$ as $X$.

\end{enumerate}
\end{minipage}}
\vspace{0.5cm}

\thm{alg-1/3}{{\rm Algorithm VC-IC[$k/3$]} finds, with probability 
at least $2.1166^{-k}$, a $k$-vertex cover of $G$, if it exists, in $O(kn)$ time,
and therefore solves {\sc Vertex Cover} in $O(2.1166^kkn)$ expected time.
}
\pf{Let $G$ be a graph with a $k$-vertex cover $X$.
By \refl{smaller-cert} for $d = 3$, $G$ contains an indirect certificate $\chi$ with at most $k/3$ vertices.
Therefore Step 1 produces a valid $(\chi, X)$-partition $(V_r, V_b)$
with probability at least $(1/4)^{k/3}(3/4)^k > 2.1166^{-k}$,
implying that the expected number of red-blue colourings to produce 
a valid $(\chi, X)$-partition is $O(2.1166^k)$.

For a valid $(\chi, X)$-partition, we have $\chi \subseteq V_r$ and $X \subseteq V_b$.
Since $X$ cover all edges, we see that $V_r$ is an independent set and hence $N(V_r) \subseteq X$.
Now by the choice of $\chi$, no vertex in $V_3(G - N[\chi])$ resides in $V - X$
and hence $V_3(G - N[\chi]) \subseteq X$, which implies $V_3(G - N[V_r]) \subseteq X$ 
as $V_3(G - N[V_r])$ is an induced subgraph of $V_3(G - N[\chi])$.
Finally $G^* = G - (N[V_r] \cup V_3(G - N[V_r]))$ is a graph of maximum degree at most two, 
i.e., a disjoint union of paths and cycles.
Therefore we can easily compute in linear time a minimum vertex cover $X^*$ of $G^*$.
Using the same arguments in the proof of \refl{smaller-cert}, 
we see that $X^*$, $N(V_r)$, and  $V_3(G - N[V_r])$ together form a minimum vertex cover of $G$.
}

\subsection{Semi-random partitions}

As shown earlier, the existence of indirect certificates for vertex covers
enables us to obtain randomized FPT algorithm based on random partition of vertices.
Here we further attest the surprising usefulness of indirect certificates
by presenting another very simple algorithm for vertex covers
based on random selection.

First we note that in Algorithm VC-IC[$k$], the colour of a vertex is totally 
independent of colours of other vertices, and the order we process vertices is immaterial.
This is in a sense wasteful as a red vertex actually forces all its neighbours 
to be blue in order to obtain a valid $(\chi, X)$-partition.
We now incorporate this forcing step into our algorithm to make the colouring stage 
semi-random, which increases the chance of obtaining a valid $(X,\chi)$-partition 
and hence the chance of success for the algorithm.

\vspace{0.5cm}
\fbox{\begin{minipage}[c]{0.9\textwidth}
\noindent Procedure Semi-Random-Partition\\
Input: Graph $G = (V,E)$.\\
Output: Red-blue partition $(V_r,V_b)$ of $V$.\\

Repeat the following until all vertices of $G$ are coloured:
Randomly choose an uncoloured vertex $v$, colour it red or blue with probability $p$
for red and probability $1-p$ for blue, and colour all neighbours of $v$ blue if $v$
is coloured red.
\end{minipage}}
\vspace{0.5cm}

We remark that the above procedure does not recolour neighbours of $v$ 
as they are either uncoloured or blue when the procedure colours $v$.
To see this, we note that if any vertex $u \in N(v)$ is red,
then $u$ is coloured red before $v$, which would have forced $v$ blue.

\lem{prob}{
Let $p$ be the probability that a vertex is coloured red.
For any vertex $v$ of degree $d$, the probability $P(v)$ that {\rm Semi-Random-Partition}
colours $v$ red and all vertices in $N(v)$ blue is at least
\[\frac{1-(1-p)^{d+1}}{d+1}\]
for $p < 1$ and $1/(d+1)$ for $p=1$.
}
\pf{First we note that for a random permutation of $N[v]$, vertex $v$
has probability $1/(d+1)$ to be in position $i$ for any $0 \le i \le d$.
For $v$ in position $i$, the $i$ vertices $v'$ of $N(v)$ before $v$ need to be blue
in order for $v$ to be red, which forces the remaining vertices in $N(v)$ to receive blue.
This gives us success probability of $(1-p)^ip$ for $v$ in position $i$, as each
vertex $v'$ has probability at least $1-p$ to be blue (note that $v'$ may have been
forced to be blue before it is processed).
Therefore
\[ P(v) \ge \frac{1}{d+1}\sum_{i=0}^d (1-p)^ip,\]
which yields the closed form in the lemma for $p < 1$ and $1/(d+1)$ for $p=1$.
}

We can replace Step 1 in Algorithm VC-IC[$k$] by Semi-Random-Partition
to obtain a new algorithm that finds in linear time a $k$-vertex cover of $G$, 
if it exists, with probability at least $(3/8)^{-k} > 2.667^{-k}$.
In fact we can fine-tune the probability $p$ to maximize the success probability of
our algorithm, and it turns out that the optimal value is $p = 1$.
This is surprising, as it suggests that there is no need for randomness in a red-blue colouring,
and it is the ordering of vertices that determines the success probability of our algorithm. 
Indeed, this perspective gives us another unexpected algorithm based on random selection:
{\em Randomly choose a vertex and declare it to be not in solution}.

\vspace{0.5cm}
\fbox{\begin{minipage}[c]{0.9\textwidth}
\noindent Algorithm VC-SRP\\
\noindent Input: A graph $G = (V,E)$. \\
\noindent Output: A $k$-vertex cover $X$ of $G$, if it exists.

\begin{enumerate}
\item Repeat the following until all vertices are coloured:
Randomly and uniformly choose an uncoloured vertex $v$,
colour $v$ red and all neighbours of $v$ blue to form a $(V_r, V_b)$-partition of $V$.
\item Output $N(V_r)$ as $X$.
\end{enumerate}
\end{minipage}}
\vspace{0.5cm}

We now use \refl{vc-cert} and \refl{prob} to prove the correctness and analyze 
the success probability of the above algorithm.
For an ordering of $V$, we write $u \prec v$ if vertex $u$ appears before vertex $v$ in the ordering.

\thm{vc-srp}{{\rm Algorithm VC-SRP} finds, with probability at least $2^{-k}$,
a $k$-vertex cover of $G$, if it exists, in $O(kn)$ time,
and therefore solves {\sc Vertex Cover} in $O(2^kkn)$ expected time.
}
\pf{As shown in the proof of \reft{vc-k}, $N(V_r)$ is a $k$-vertex cover
once $(V_r,V_b)$ is a valid $(\chi, X)$-partition.
To establish the theorem, we show that Semi-Random-Partition
has probability at least $2^{-k}$ to produce a valid $(\chi,X)$-partition $(V_r,V_b)$.
First we note that the probability of producing a valid $(\chi,X)$-partition 
only depends on the subgraph consisting of cut edges $[\chi, X]$.
Let $G^*$ be such a subgraph with fewest edges and minimum probability $p(G^*)$ 
of producing a valid $(\chi,X)$-partition for $|X| =k$, and
we examine the structure of $G^*$ to determine $p(G^*)$.

If $G^*$ has two vertices $u, v \in\chi$ sharing a common neighbour $w$,
let $G' = G^* -uw$ and we show that $p(G') \le p(G^*)$.
For this purpose, it suffices to consider a bad vertex ordering of $G^*$ that
colours $u$ improperly (i.e., blue) as a consequence of colouring $w$ red.
For this to happen, we have $w \prec u$ and $w$ is uncoloured when it is processed.
Therefore we also have $w \prec v$ since otherwise $w$ is forced to be blue when
$v$ is processed, implying that $v$ is also forced to be blue when $w$ is processed.
Therefore the ordering is also a bad vertex ordering for $G'$, 
which implies that $p(G') \le p(G^*)$, contradicting the choice of $G^*$.

Therefore $G^*$ consists of a disjoint union of $t$ stars $S_i$ with vertices $\chi$ as
centres.
By \refl{prob}, each $S_i$ has probability $1/s_i$ to be properly coloured
if $S_i$ contains $s_i$ vertices.
Since the colouring of stars $S_i$ is independent, we have
\[ p(G^*) = \prod_{1 \le i \le t}\frac{1}{s_i}  \]
with $\sum_{1\le i \le t} (s_i - 1) \le k$.
It is easy to see that $p(G^*)$ achieves its minimum value $2^{-k}$ when every $s_i = 2$,
i.e., $G^*$ is the disjoint union of $k$ edges.
Therefore Semi-Random-Partition has probability at least $2^{-k}$ to produce 
a valid $(\chi,X)$-partition $(V_r,V_b)$, and hence the expected number of
repeated runs of the algorithm to produce a valid $(\chi,X)$-partition is $O(2^k)$.}

We remark that the success probability of {\rm Algorithm VC-SRP} can be improved to
$1.6633^{-k}$ by using indirect certificates of size $k/3$.

\section{Minimum Vertex Covers}

The iterative compression method of Reed, Smith, and Vetta~\cite{Reed}
is a useful and powerful tool for designing FPT algorithms.
The key idea of the method is to compress a given solution to a smaller one
if it can.

We will use the iterative compression method to obtain two FPT algorithms
for {\sc Vertex Cover}.
For this purpose, we first turn our attention to determine whether a
given vertex cover of $G = (V, E)$ is a minimum vertex cover.

Let $X$ be a set of vertices in $G$.
For a subset $I \subseteq X$, the {\em outside neighbourhood} of $I$ (w.r.t. $X$),
denoted $N^*(I)$, consists of vertices in $V - X$ that are adjacent 
to some vertices in $I$, i.e., $N^*(I) = \bigcup_{v \in I} N(v) - X$.
The following characterization of a minimum vertex cover can well be a 
very old folklore in graph theory, and was observed by the author in 2005
when he was desperately looking for simple examples to teach the iterative 
compression method.

\lem{minvc}{A vertex cover $X$ of a graph $G = (V, E)$ is a minimum vertex cover iff
for every independent set $I$ of $G[X]$, $|N^*(I)| \ge |I|$.}
\pf{
If $G[X]$ contains an independent set $I$ with $|N^*(I)| < |I|$,
then $(X - I) \cup N^*(I)$ is a vertex cover of $G$ smaller than $X$
as all edges covered by $I$ are covered by $X - I$ and $N^*(I)$.
Conversely, suppose that $G$ has a smaller vertex cover $X'$.
Let $I = X - X'$.
Then $I$ is an independent set of $G[X]$, and $N^*(I) \subseteq X' - X$.
Since $|X'| < |X|$, we have $|X' - X| < |X - X'|$ and hence $|N^*(I)| < |I|$.
}

The above lemma enables us to determine in FPT-time whether a $(k+1)$-vertex cover 
of $G$ is a minimum vertex cover, and obtain a smaller one when it is not.
The following compression routine takes $O(2^kkn)$ time as at most 
$2^{k+1}$ subsets of $X$ need to be examined in the worst case:

\vspace{0.5cm}
\fbox{\begin{minipage}[c]{0.9\textwidth}
Consider each independent set $I$ in a $(k+1)$-vertex cover $X$, and check if
$|N^*(I)| < |I|$. 
If so we get a smaller vertex cover $(X - I) \cup N^*(I)$, 
otherwise $G$ has no $k$-vertex cover.
\end{minipage}}
\vspace{0.5cm}

As for the question of how to obtain a $(k+1)$-vertex cover in the first place,
there are at least three different ways:

\begin{enumerate}
\item {\em Recursively}: Arbitrarily choose a vertex $v$ in $G$ and recursively
solve the problem for $G-v$.
If $G-v$ has no solution then neither has $G$.
Otherwise, we obtain a $k$-vertex cover $X$ of $G - v$, which yields a $(k+1)$-vertex cover
$X \cup \{v\}$ for $G$.
We need to compress $O(n)$ times, which results in an $O(2^kkn^2)$-time algorithm.

\item {\em Iteratively}: Order vertices as $v_1,\dots,v_n$
and let $G_i = G[\{v_1,\dots,v_i\}]$.
Then a $k$-vertex cover $X$ of $G_i$ yields a $(k+1)$-vertex cover
$X \cup \{v_{i+1}\}$ of $G_{i+1}$, which is then compressed into a $k$-vertex cover.
Again, we get an $O(2^kkn^2)$-time algorithm.

\item  {\em Approximation}: Use a 2-approximation algorithm for vertex covers
to find a $k'$-vertex cover of $G$ first.
If $k' > 2k$ then the answer is no, otherwise we compress solutions
at most $k$ times, which gives us an $O(4^kk^2n)$-time algorithm.
\end{enumerate}

Algorithmically, it is not efficient to consider all possible independent
sets $I$ inside $X$, and in fact it is sufficient to consider maximal
independent sets only.

\lem{minvc-maxis}{A vertex cover $X$ of a graph $G = (V, E)$ is a minimum vertex cover 
iff for every maximal independent set $I$ in $X$, $I$ is a minimum vertex cover 
of the induced subgraph $G[I \cup N^*(I)]$.
}
\pf{
If there is a maximal independent set $I \subseteq X$ such that the induced subgraph
$G[I \cup N^*(I)]$ admits a vertex cover $X'$ of size smaller than $I$,
then $(X - I) \cup X'$ is a smaller vertex cover of $G$ as $X - I$ covers 
all edges outside $G[I \cup N^*(I)]$.

Conversely, suppose that $G$ has a vertex cover $X'$ smaller than $X$.
By \refl{minvc}, there is an independent set $I' \subseteq X$ satisfying
$|N^*(I')| < |I'|$.
Let $I \subseteq X$ be a maximal independent set containing $I'$.
Then $G[I \cup N^*(I)]$ is a bipartite graph as $N^*(I) \subseteq V -X $ is  
an independent set, and hence admits $(I - I') \cup N^*(I')$ as a vertex cover,
which is smaller than $I$ as $|N^*(I')| < |I'|$.
}

With \refl{minvc-maxis} in hand, we obtain a different compression routine
which examines fewer subsets of $X$:

\vspace{0.5cm}
\fbox{\begin{minipage}[c]{0.9\textwidth}
Consider each maximal independent set $I$ in a $(k+1)$-vertex cover $X$, and check if
$G[I \cup N^*(I)]$ has a vertex cover $X'$ smaller than $I$.
If so we obtain a smaller vertex cover $(X - I) \cup X'$, otherwise $G$ has no $k$-vertex cover.
\end{minipage}}
\vspace{0.5cm}

The above compression routine takes $O(3^{k/3}kn^{1.5})$ time as $G[X]$ can contain
$O(3^{k/3})$ maximal independent sets in the worst case~\cite{Moon}, which can be listed 
in time $O(3^{k/3}(m + n))$~\cite{Tomita}, and it takes 
$O(m\sqrt{n})$ time to find a minimum vertex cover in a bipartite graph~\cite{Karp}.
Recall that $m = O(kn)$ for graphs with $k$-vertex covers.

\section{Colourful vertex covers}

Finally we turn to the colour coding method of Alon, Yuster and Zwick~\cite{Alon}
for {\sc Vertex Cover}.
The basic idea of this novel method is to use $k$ colours to colour elements
randomly and then try to find a colourful $k$-solution, i.e., a $k$-solution
whose elements are in distinct colours.
If we can find a colourful $k$-solution in FPT time, then we can find
a $k$-solution in FPT time with probability $k!/k^k > e^{-k}$.
The algorithm can be derandomized by a family of perfect hash functions.

The colour coding method is typically useful for parameterized problems whose
$k$-solutions either have a nice structure (e.g., linear, cyclic, tree) or
can be partitioned into independent structures of constant size (e.g., triangle packing).
Although {\sc Vertex Cover} does not seem to possess such properties,
we will show that the method is applicable to the problem.
In fact, we will present a linear algorithm 
for the colourful version of {\sc Vertex Cover}.

Let $G = (V, E; f)$ be a vertex coloured graph with $f: V \rightarrow \{1,\dots,k\}$.
We use $V_i$ to denote the set of vertices with colour $i$, and call each $V_i$
a {\em colour class}.
A vertex cover $X$ of $G$ is {\em colourful} if all vertices in $X$ have distinct 
colours, i.e., $X$ contains at most one vertex from each colour class, and we
will design a linear algorithm for the following problem:

\begin{quote}
{\sc Colourful Vertex Cover}\\
{\sc Instance}: Vertex coloured graph $G$ with colours in $\{1,\dots,k\}$.\\
{\sc Question}: Does $G$ contain a colourful vertex cover?
\end{quote}

First we note that the problem is no easier than {\sc 2Sat} as we can 
reduce {\sc 2Sat} to our problem in linear time: For an arbitrary instance 
$(U, C)$ of {\sc 2Sat}, we construct a vertex coloured graph $G$ by creating,
for each Boolean variable $u_i \in U$, two vertices $u_i$ and $\overline{u}_i$ 
with colour $i$ and edge $u_i\overline{u}_i$; and adding, for each binary clause $\{x, y\} \in C$, 
an edge between vertices $x$ and $y$.

Inspired by the above connection with {\sc 2SAT}, we first reduce 
{\sc Colourful Vertex Cover} to {\sc 2SAT} to obtain a quadratic algorithm,
and will improve it to a linear algorithm later.
For the purpose of a quadratic algorithm, we construct a Boolean formula 
$\Phi(G)$ for $G$ as follows:
\begin{enumerate}
\item For each vertex $v$, introduce a Boolean variable $x_v$.
\item For edge set $E$, let $\Phi(E) = \bigwedge_{uv \in E} x_u \vee x_v$.
\item For each colour class $V_i$, let
$\Phi(V_i) = \bigwedge_{u, v \in V_i \mbox{ and } u \not= v} \overline{x_u \wedge x_v}$.
\item Set $\Phi(G) = \Phi(E) \bigwedge_{i = 1}^k \Phi(V_i)$.
\end{enumerate}

\thm{2sat-colourful}{
A vertex coloured graph $G$ admits a colourful vertex cover iff 
its corresponding formula $\Phi(G)$ is satisfiable.
}
\pf{
Clearly {\sc Colourful Vertex Cover} is equivalence to the following 
integer linear programming: For all $v \in V$, find $x_v \in \{0, 1\}$ 
to satisfy
\[ x_u + x_v \ge 1 \mbox{ for each edge $uv$ of $G$, and}\]
\[ \sum_{v \in V_i} x_v \le 1 \mbox{ for each colour class $V_i$ of $G$}. \]
Note that a vertex $v$ belongs to a colourful vertex cover iff $x_v = 1$.

Now for any $x, y \in \{0, 1\}$, $x + y \ge 1$ is equivalent to $x \vee y$
when we also interpret $x$ and $y$ as Boolean variables.
Similarly, $x + y \le 1$ is equivalent to $\overline{x \wedge y}$.
Furthermore, for each colour class $V_i$, $\sum_{v \in V_i} x_v \le 1$
is equivalent to $x_u + x_v \le 1$ for all distinct $u, v \in V_i$.
It follows that $\Phi(G)$ is satisfiable iff the above integer linear programming 
has a solution, and equivalently $G$ admits a colourful vertex cover. 
}

The above theorem enables us to solve {\sc Colourful Vertex Cover} in $O(n^2)$ time
as $\Phi(G)$ contains $O(n^2)$ binary clauses 
(note that $\overline{x \wedge y} = \overline{x} \vee \overline{y}$), 
and {\sc 2Sat} is solvable in linear time~\cite{PS}.
However, it is unlikely that we can obtain a linear algorithm by reduction 
to 2SAT as the number of binary clauses seems quadratic for any such reduction.
To obtain a linear algorithm, we solve our problem directly by following the idea 
of limited backtracking for 2SAT.

The key to our linear algorithm is the following forcing properties for edges and colour classes.
Recall that for a vertex $v$, $v$ belongs to a colourful vertex cover iff $x_v = 1$.

\begin{enumerate}
\item For an edge, if one end has value 0 then the other end is forced to 1. 
\item For a colour class, if one vertex inside it has value 1 then 
all other vertices in the class are forced to 0.
\end{enumerate}
Therefore the value of a vertex $v$ may force values on other vertices, which may cause 
a chain reaction to force values on more vertices.
Of course, forcing may cause a conflict of values for some vertices.
For any $b \in \{0, 1\}$ and vertex $v$, we define $F_b(v)$ to be the set of vertices
that are forced to receive a value when $x_v = b$.
Assign $F_b(v) = \es$ if $x_v = b$ causes a conflict of value for any forced vertex,
which indicates $x_v \not= b$.
The following lemma allows us to apply self-reduction to $G$ whenever $F_b(v) \not= \es$
for any $b \in \{0, 1\}$.

\lem{reduction}{
For any $b \in \{0, 1\}$ and vertex $v$, $G$ admits a colourful vertex cover iff $G - F_b(v)$ admits one.
}
\pf{
Let $G' = G - F_b(v)$ and $X$ a colourful vertex cover of $G$. 
Since edges of $G'$ can be covered by vertices in $G'$ only,
$X$ restricted to $G'$ is clearly a colourful vertex cover of $G'$.

Conversely, suppose that $G'$ admits a colourful vertex cover $X'$ and let 
$F' \subseteq F_b(v)$ be vertices with value 1.
Then $F'$ is colourful and covers all edges of $G$ outside $G'$, 
and hence $F' \cup X'$ is a vertex cover of $G$.
For each colour class $V'_i$ in $G'$, $V_i - V'_i$ contains no vertex of $F'$
by the definition of $F_b(v)$. 
Therefore $F' \cup X'$ is colourful.
}

The above lemma naturally leads us to the following algorithm which, for clarity,
is presented as a parallel algorithm.
In our algorithm,  we use two processors $P_0$ and $P_1$ to compute 
$F_0(v)$ and $F_1(v)$ independently for a vertex $v$.
For efficiency, we stop processor $P_b$ each time when processor 
$P_{\overline{b}}$ has obtained a nonempty $F_{\overline{b}}(v)$.
Our algorithm can be converted into a sequential one 
by the standard dovetailing technique.

\vspace{0.5cm}
\fbox{\begin{minipage}[c]{0.9\textwidth}
\begin{tabbing}
Algorithm Colourful-VC \\
Input: A vertex coloured graph $G = (V, E; f)$ with $f: V \rightarrow \{1,\dots,k\}$.\\
Output: A colourful vertex cover $X$ of $G$. \\ \\

{\bf while} $G$ contains a vertex $v$ with no value {\bf do} the following in parallel:\\
Processor $P_0$: \= compute $F_0(v)$; \\
\> {\bf if} $F_0(v) \not=\es$ {\bf then} \= {\bf stop $P_1$}, \\
\> \> assign values to $F_0(v)$ accordingly, \\ 
\> \> and $G \leftarrow G - F_0(v)$; \\
Processor $P_1$: compute $F_1(v)$; \\
\> {\bf if} $F_1(v) \not=\es$ {\bf then} \= {\bf stop $P_0$}, \\
\> \> assign values to $F_1(v)$ accordingly, \\
\> \> and $G \leftarrow G - F_1(v)$; \\
{\bf if} both $F_0(v)$ and $F_1(v)$ are empty {\bf then} return ``No solution'' and {\bf halt}; \\
{\bf end while}; \\
Return all vertices with value 1 as $X$. 
\end{tabbing}
\end{minipage}}
\vspace{0.5cm}

We can compute $F_b(v)$ by BFS with one of the following two actions
in extending the current vertex $u$ to other vertices.
The action depends on the value of the current vertex $u$.

\begin{tabbing}
{\bf Case $x_u = 0$}: {\bf for} \= each edge $uu'$ {\bf do} \\
		\> {\bf if} $x_{u'} = 0$ \=  {\bf then} $F_b(v) \leftarrow \es$ and {\bf stop} \\
		\> \> {\bf else} $x_{u'} \leftarrow 1$.\\

{\bf Case $x_u = 1$}: {\bf for} every other vertex $u'$ in the colour class containing $u$ {\bf do} \\
		\> {\bf if} $x_{u'} = 1$ {\bf then} $F_b(v) \leftarrow \es$ and {\bf stop} \\
		\> \> {\bf else} $x_{u'} \leftarrow 0$.
\end{tabbing}

\thm{time-colourful}{
For any vertex coloured graph $G$ with $k$ colours, it takes $O(kn)$ time to find 
a colourful vertex cover in $G$, if it exists.
}
\pf{
The correctness of algorithm {\rm Colourful-VC} follows from \refl{reduction}.
If both $F_0(v)$ and $F_1(v)$ are empty, then the algorithm clearly halts in time 
$O(m+n)$ as it just performs BFS twice. 
Otherwise one of them is not empty and we may assume that, without loss of generality,
$F_0(v)$ is not empty and processor $P_0$ finishes before $P_1$.
Let $T(G)$ denote the time of the algorithm for processing graph $G$,
and we have the following recurrence:
\[ T(G) = T(G - F_0(v)) + O(\mbox{the number of edges covered by $F_0(v)$}) \]
since the latter term is the time spent by processor $P_0$ to compute $F_0(v)$. 
It is obvious by induction that $T(G) = O(m + n) = O(kn)$ as $m =O(kn)$ for
graphs with $k$-vertex covers.
}

The above theorem implies that the colour coding method can be used to 
solve {\sc Vertex Cover} in FPT time.
We remark that the problem becomes intractable if we wish to minimize 
the size of $X$, or allow $X$ to contain at most two vertices from each colour class.

\section{Concluding remarks}

We have presented half a dozen new and simple FPT algorithms for the classical 
{\sc Vertex Cover} by using iterative compression, colour coding, 
and indirect certificating.
These algorithms explore structural properties of the problem, which  
deepens our understanding of the problem.
In particular, the existence of indirect certificates of size $k/3$ for {\sc Vertex
Cover} is quite interesting and surprising.
We hope that ideas in the paper will be useful in designing FPT algorithms 
for other problems --- although we have to do exhaustive search one way or another, 
it is fun and challenging to do such bad things in clever ways.

The indirect certificating technique seems quite potential for designing FPT algorithm, 
though several problems only are solved by the method so far.
The existence of small indirect certificates seems to have connection 
with kernelization: we always have a small indirect certificate if a kernel 
is uniquely determined by a kernelization algorithm.
It is worthwhile to look at a problem from the perspective of small indirect certificates, 
which may bring us new insight into the problem.

\pro{kernel}{Find sufficient conditions for kernels to be useful for small indirect
certificates.}

Of course, indirect certificates seem interesting in their own right.
In particular, certificates of size a fraction of solution size have both theoretical 
and practical values.
We note that certificates of size $k/3$ for {\sc Vertex Cover} implies that 
if there is a polynomial reduction from a problem $\Pi$ to {\sc Vertex Cover} that
preserves solution size $k$, then $\Pi$ also admits certificates of size $k/3$.
Using this reduction approach, we can infer that the following two NP-hard
problems both have certificates of size at most $k/3$: 
find a weight-$k$ truth assignment for {\sc 2Sat}~\cite{MRS},
and delete $k$ edges to destroy all alternating paths
in an edge coloured graph~\cite{CL}.

\pro{sic}{What kinds of problems have certificates whose size is a fraction of solution size?}

In connection with FPT algorithms, it is actually not hard to propose 
various small indirect certificates $\chi$.  
The pressing question is how to make use of $\chi$ to find $X$, and the main obstacle 
seem to be noises caused by red elements not from $\chi$.

\pro{noise}{For a valid $(\chi, X)$-partition, under what conditions can we 
find a $k$-solution in FPT time?}

Coming back to {\sc Vertex Cover}, we can reduce the size of certificates from 
$k/3$ to $k/3 - c$ for any constant $c$, but it seems unlike to shave off $\log k$
from $k/3$ as in the $2k - c\log k$ kernelization~\cite{Lampis} of the problem.

\pro{vc}{Does {\sc Vertex Cover} have certificates of size $k/d$ for some $d > 3$?}

Finally, in echo to the algorithm that starts the paper, we conclude with 
the following algorithm that uses random orientation to solve with probability at least
$2^{-k}$ the related {\sc Partial Vertex Cover} problem: Find fewest vertices 
in a graph $G$ to cover at least $k$ edges.

\vspace{0.5cm}
\fbox{\begin{minipage}[c]{0.9\textwidth}
Randomly orient each edge to obtain a digraph $\vec{G}$ from $G$. 
Order vertices $\vec{G}$ nonincreasingly by their out-degrees, and choose vertices 
until the sum of their out-degrees is at least $k$.
\end{minipage}}
\vspace{0.5cm}


\bibliography{ref}
{}

\bibliographystyle{plain}

\end{document}